\documentstyle[preprint,pra,aps]{revtex}
\begin{document}

\draft

\title{Condensate formation in a Bose gas}

\author{H.T.C. Stoof}
\address{Department of Physics, University of Illinois at Urbana-Champaign,
	 1110 West Green Street, Urbana, Illinois 61801, U.S.A. and \\
	 Department of Theoretical Physics, Eindhoven University of
	 Technology, P.O. Box 513, \\ 5600 MB Eindhoven, The Netherlands
	  \cite{address}}

\maketitle

\begin{abstract}
Using magnetically trapped atomic hydrogen as an example, we investigate the
prospects of achieving Bose-Einstein condensation in a dilute Bose gas. We
show that, if the gas is quenched sufficiently far into the critical region
of the phase transition, the typical time scale for the nucleation of the
condensate density is short and of $O(\hbar/k_{B}T_{c})$. As a result we find
that thermalizing elastic collisions act as a bottleneck for the formation of
the condensate. In the case of doubly-polarized atomic hydrogen these occur
much more frequently than the inelastic collisions leading to decay and we
are lead to the conclusion that Bose-Einstein condensation can indeed be
achieved within the lifetime of the gas.
\end{abstract}


\section{INTRODUCTION}

In the last few years it has been clearly demonstrated that not only charged
ions but also neutral atoms can be conveniently trapped and cooled by means
of electro-magnetic fields. Although the physics of the various ingenious
scenarios developed to accomplish this is already interesting in itself
\cite{chu}, the opportunities offered by an atomic gas sample at very low
temperatures are exciting too. Examples in this respect are the performance
of high-precision spectroscopy, the search for a violation of CP invariance
by measuring the electric dipole moment of atomic cesium \cite{bern}, the
construction of an improved time standard based on an atomic fountain
\cite{clairon}, and the achievement of Bose-Einstein condensation in a
weakly-interacting gas.

In particular the last objective is an important motivation for studying
cold atomic gases and has been pursued most vigorously with atomic hydrogen
\cite{mit,adam}. However, it was recently proposed that also the alkali-metal
vapors cesium \cite{wieman} and lithium \cite{hulet} are suitable candidates
for the achievement of Bose-Einstein condensation. We will nevertheless
concentrate here on atomic hydrogen, because it still seems to be the most
promising system for the observation of the phase transition in the near
future. Moreover, it has the advantage that the atomic interaction potential
is known to a high degree of accuracy. As a result we can have confidence in
the fact that the scattering length is positive, which is required for the
condensation to take place in the gaseous phase \cite{nega}, and small enough
to rigorously justify the approximations made in the following for the
typical temperatures ($T \simeq 10 \: \mu K$) and densities
($n \simeq 1 \cdot 10^{14} \: cm^{-3}$) envisaged in the experiments.

Due to the spin of the electron and the proton, the 1s-hyperfine manifold
of atomic hydrogen consists of four states which are in order of increasing
energy denoted by $|a \rangle$, $|b \rangle$, $|c \rangle$, and
$|d \rangle$, respectively. Only the $|c \rangle$ and $|d \rangle$ states
can be trapped in a static magnetic trap, because in a magnetic field they
have predominantly an electron spin-up component and are therefore
low-field seeking \cite{review}. Furthermore, if we load a trap with atoms
in these two hyperfine states, the $|c \rangle$ state is rapidly depopulated
as a result of the much larger probability for collisional relaxation to the
high-field seeking $|a \rangle$ and $|b \rangle$ states which are
expelled from the trap. In this manner the system polarizes spontaneously and
we obtain a gas of $|d \rangle$-state atoms, known as doubly spin-polarized
atomic hydrogen since both the electron as well as the proton spin are
directed along the magnetic field. Unfortunately, such a doubly-polarized
hydrogen gas still decays due to the dipole interaction between the magnetic
moments of the atoms. Although the time scale $\tau_{inel}$ for this decay is
much longer than the time scale for the depopulation of the $|c \rangle$
state mentioned above, it nevertheless limits the lifetime of the gas sample
to the order of seconds for the densities of interest \cite{rates}.

Having filled the trap with doubly-polarized atoms, we must subsequently
lower the temperature of the gas to accomplish Bose-Einstein condensation.
At present it is believed that the most convenient way to achieve this is
by means of conventional \cite{doyle,luiten} or light-induced \cite{setija}
evaporative cooling. In both cases the idea is to remove, by lowering the
well-depth or by photon absorption in the perimeter, the most energetic
particles from the trap and thus to create momentarily a highly
nonequilibrium energy distribution that will evolve into a new equilibrium
distribution at a lower temperature. According to the quantum Boltzmann
equation describing this process, a typical time scale for the evolution is
the average time between two elastic collisions
$\tau_{el} = 1/n \langle v \sigma \rangle$, with $\langle v \sigma \rangle$
the thermal average of the relative velocity $v$ of two colliding atoms times
their elastic cross section $\sigma$. Clearly, $\tau_{el}$ must be small
compared to $\tau_{inel}$ to ensure that thermal equilibrium is achieved
within the lifetime of the system. As a result, the minimum temperature that
can be reached by evaporative cooling is about $1\: \mu K$ and indeed below
the critical temperature of atomic hydrogen at a density of
$1 \cdot 10^{14} \: cm^{-3}$.

The previous discussion appears to indicate that a typical time scale for
the formation of the condensate is given by $\tau_{el}$. However, this is not
correct because simple phase-space arguments show that a kinetic equation
cannot lead to a macroscopic occupation of the one-particle ground state:
Considering a homogeneous system of $N$ bosons in a volume $V$, we find from
the Boltzmann equation that the production rate of the condensate fraction is
\begin{equation}
\frac{d}{dt} \left. \frac{N_{\vec 0}}{N} \: \right|_{in}
			= C \frac{\langle v \sigma \rangle}{V}
			    (1+N_{\vec 0}) \:,
\end{equation}
where $N_{\vec 0}$ is the number of particles in the zero-momentum state and
$C$ is a constant of $O(1)$. Hence, in the thermodynamic limit
($N,V \rightarrow \infty$ in such a way that their ratio $n=N/V$ remains
fixed) a nonzero production rate is only possible if a condensate already
exists \cite{levich} and we are forced to conclude that Bose-Einstein
condensation cannot be achieved by evaporative cooling of the gas.

\section{NUCLEATION}

In the above argument we have only considered the effect of two-body
collisions. It is therefore legitimate to suggest that perhaps three or
more body collisions are required for the formation of the condensate,
even though they are very improbable in a dilute gas \cite{snoke}. However,
we can easily show that the same argument also applies to these
processes: For a $m$-body collision that produces one particle with zero
momentum we have $2m-2$ independent momentum summations, leading to a
factor of $V^{2m-2}$. Moreover, the transition matrix element is
proportional to $V \cdot V^{-m}$ due to the integration over the
center-of-mass coordinate and the normalization of the initial and final
state wave functions, respectively. In total the production rate for the
condensate fraction is thus proportional to
$V^{2m-2} (V^{1-m})^{2} V^{-1} (1+N_{\vec 0})$ or $V^{-1} (1+N_{\vec 0})$,
which again vanishes in the thermodynamic limit if there is no nucleus of
the condensed phase. As expected, the contributions from collisions that
produce more than one zero-momentum particle have additional factors of
$V^{-1}$ and vanish even more rapidly if $V \rightarrow \infty$.

Clearly, we have arrived at a nucleation problem for the achievement of
Bose-Einstein condensation which seriously endangers the success of future
experiments. Fortunately, we suspect that the line of reasoning presented
above is not completely rigorous because otherwise it implies that also
liquid helium cannot become superfluid, in evident disagreement with our
experience. Indeed, by using a kinetic equation to discuss the time evolution
of the gas we have in effect neglected the buildup of coherence which is
crucial for the formation of the condensate. Our previous argument therefore
only shows that by means of evaporative cooling the gas is quenched into the
critical region on a time scale $\tau_{el}$, not that Bose-Einstein
condensation is impossible. To discuss that point we need a different
approach that accurately describes the time evolution of the system after the
quenching by taking the buildup of coherence into account exactly. Such a
nonequilibrium approach was recently developed on the basis of the Keldysh
formalism and can, in the case of a dilute Bose gas, be seen as a
generalization of the Landau theory of second-order phase transitions
\cite{stoof}. As a consequence it is useful to consider the Landau theory
first. This leads to a better understanding of the more complicated
nonequilibrium theory and ultimately of the physics involved in the
nucleation of Bose-Einstein condensation.

\subsection{Landau theory}

As an introduction to the Landau theory of second-order phase transitions we
use the example of a ferromagnetic material \cite{landau}. To be more
specific we consider a cubic lattice with spins $\vec{S}_{i}$ at the sites
$\{i\}$. The Hamiltonian is taken to be
\begin{equation}
\label{energy}
H = -J \sum_{\langle i,j \rangle} \vec{S}_{i} \cdot \vec{S}_{j} \:,
\end{equation}
where $J$ is the exchange energy and the sum is only over nearest neighbors.
For further convenience we also introduce the magnetization
\begin{equation}
\vec{M} = \frac{1}{V} \sum_{i} \vec{S}_{i} \:.
\end{equation}

Physically it is clear that this model has a phase transition at a critical
temperature $T_{c}$ of $O(J/k_{B})$. Above the critical temperature the
thermal fluctuations randomize the direction of the spins and the system is
in a disordered (paramagnetic) state having a vanishing average magnetization
${\langle \vec{M} \rangle}_{eq}$. However, below the critical temperature the
thermal fluctuations are not large enough to overcome the directional effect
of the Hamiltonian and the spins favor an ordered (ferromagnetic) state with
${\langle \vec{M} \rangle}_{eq} \neq \vec{0}$. The different phases of the
material are thus conveniently characterized by the average magnetization,
which for this reason is known as the order parameter of the ferromagnetic
phase transition.

In the phenomenological approach put forward by Landau the above mentioned
temperature dependence of the equilibrium order parameter
${\langle \vec{M} \rangle}_{eq}$ is reproduced by anticipating that the
free-energy density of the system at a fixed but not necessarily equilibrium
value of the average magnetization has the following expansion
\begin{equation}
\label{fenergy}
f(\langle \vec{M} \rangle,T) \simeq f(\vec{0},T) +
			     \alpha(T) {\langle \vec{M} \rangle}^{2} +
		    \frac{\beta(T)}{2} {\langle \vec{M} \rangle}^{4}
\end{equation}
for small values of $\langle \vec{M} \rangle$, and that the coefficients of
this expansion behave near the critical temperature as
\begin{equation}
\alpha(T) \simeq \alpha_{0} \left( \frac{T}{T_{c}} - 1 \right)
\end{equation}
and
\begin{equation}
\beta(T) \simeq \beta_{0} \:,
\end{equation}
respectively, with $\alpha_{0}$ and $\beta_{0}$ positive constants.

Hence, above the critical temperature $\alpha(T)$ and $\beta(T)$ are both
positive. As a result the free energy is minimal for
$\langle \vec{M} \rangle = \vec{0}$, which corresponds exactly to the
paramagnetic phase. Moreover, for temperatures below the critical one
$\alpha(T)$ is negative and the free energy is indeed minimized by a nonzero
average magnetization with magnitude $\sqrt{- \alpha(T) / \beta(T)}$. Just
below the critical temperature the latter equals
\begin{equation}
{\langle M \rangle}_{eq} \simeq \sqrt{ \frac{\alpha_{0}}{\beta_{0}}
				\left( 1 - \frac{T}{T_{c}} \right) } \;,
\end{equation}
which after substitution in Eq.\ (\ref{fenergy}) gives rise to an equilibrium
free-energy density of
\begin{equation}
f({\langle \vec{M} \rangle}_{eq},T) \simeq f(\vec{0},T) -
		  \frac{\alpha_{0}^{2}}{2\beta_{0}}
			{\left( 1 - \frac{T}{T_{c}} \right)}^{2} \:.
\end{equation}
Therefore, the second derivative $d^{2}f/dT^{2}$ is discontinuous at the
critical temperature and the phase transition is of second order according
to the Ehrenfest nomenclature.

Note that minimizing the free energy only fixes the magnitude and not the
direction of ${\langle \vec{M} \rangle}_{eq}$. This degeneracy is caused by
the fact that the Hamiltonian in Eq.\ (\ref{energy}) is symmetric under an
arbitrary rotation of all the spins $\vec{S}_{i}$. Consequently, the free
energy must be symmetric under a rotation of the average magnetization and
only even powers of $\langle \vec{M} \rangle$ can appear in its expansion
(cf.\ Eq.\ (\ref{fenergy})). Due to this behavior the ferromagnet is a good
example of a system with a spontaneously broken symmetry, i.e.\ although the
Hamiltonian is invariant under the operations of a group, its ground state
is not. In the case of a ferromagnet the symmetry group is $SO(3)$, which is
broken spontaneously below the critical temperature because the average
magnetization points in a certain direction. Which direction is chosen in
practice, depends on the surroundings of the system and in particular on
(arbitrary small) external magnetic fields that favor a specific direction.

After this summary of the Landau theory we are now in a position to
introduce two time scales which turn out to be of great importance for the
nucleation of Bose-Einstein condensation. To do so we consider the following
experiment: Imagine that we have a piece of ferromagnetic material
at some temperature $T_{1}$ above the critical temperature. Being in thermal
equilibrium the material is in the paramagnetic phase with
${\langle \vec{M} \rangle}_{eq} = \vec{0}$. We then quickly cool the
material to a new temperature $T_{2}$ below the critical temperature.
If this is done sufficiently fast, the spins will have no time to react and
we obtain a nonequilibrium situation in which the free energy has developed
a `double-well' structure but the average magnetization is still zero. This
is depicted in Fig.\ \ref{fig1}(a). In such a situation there is a typical
time scale for the relaxation of the average magnetization to its new
equilibrium value $\sqrt{- \alpha(T_{2}) / \beta(T_{2})}$, which we denote
$\tau_{coh}$.

However, in the case of magnetically trapped atomic hydrogen, the gas is
isolated from its surroundings and it is not possible to perform the cooling
stage mentioned above. As a result the gas has to develop the instability
associated with the phase transition by itself. The time scale corresponding
to this process is called $\tau_{nucl}$ and is schematically shown in
Fig.\ \ref{fig1}(b). Combining the two processes we are lead to the
following physical picture for the nucleation of Bose-Einstein
condensation. After the quench into the critical region the gas develops an
instability on the time scale $\tau_{nucl}$. On this time scale the actual
nucleation takes place and a small nucleus of the condensate is formed, which
then grows on the time scale $\tau_{coh}$ as shown in Fig.\ \ref{fig2}.
To solve the nucleation problem we are thus left with the actual
determination of these two time scales. Clearly, before this can be done we
need to know the correct order parameter of the phase transition.

\subsection{Order parameter}

Ever since the pioneering work of Bogoliubov \cite{bogol} it is well
known that the order parameter for Bose-Einstein condensation in a
weakly-interacting Bose gas is a somewhat abstract quantity, which is most
conveniently discussed by using the method of second quantization. In this
method all many-body observables are expressed in terms of the creation and
annihilation operators of a particle at position $\vec{x}$ denoted by
$\psi^{\dagger}(\vec{x})$ and $\psi(\vec{x})$, respectively \cite{fetter}.
For example, for a gas of particles with mass $m$ and a two-body interaction
potential $V(\vec{x}-\vec{x}')$ the Hamiltonian equals
\begin{equation}
\label{hamil}
H = \int d\vec{x} \: \psi^{\dagger}(\vec{x}) \frac{-\hbar^{2}\nabla^{2}}{2m}
		     \psi(\vec{x}) +
    \frac{1}{2} \int d\vec{x} \int d\vec{x}' \:
		  \psi^{\dagger}(\vec{x}) \psi^{\dagger}(\vec{x}')
		  V(\vec{x}-\vec{x}')
		  \psi(\vec{x}') \psi(\vec{x})
\end{equation}
and the total number of particles is given by
\begin{equation}
\label{number}
N = \int d\vec{x} \: \psi^{\dagger}(\vec{x}) \psi(\vec{x}) \:.
\end{equation}
The method is also particularly useful for a Bose system because the
permutation symmetry of the many-body wave function is automatically
accounted for by assuming the commutation relations
$[\psi(\vec{x}),\psi(\vec{x}')] =
	      [\psi^{\dagger}(\vec{x}),\psi^{\dagger}(\vec{x}')] = 0$ and
$[\psi(\vec{x}),\psi^{\dagger}(\vec{x}')] = \delta(\vec{x}-\vec{x}')$
between the creation and annihilation operators.

In the language of second quantization the order parameter for the dilute
Bose gas is the expectation value $\langle \psi(\vec{x}) \rangle$.
Analogous to the case of the ferromagnetic phase transition, a nonzero
value of this order parameter signals a spontaneously broken symmetry.
Here the appropriate symmetry group is $U(1)$, since the Hamiltonian of
Eq.\ (\ref{hamil}) is invariant under the transformation
$\psi(\vec{x}) \rightarrow \psi(\vec{x})e^{i\vartheta}$ and
$\psi^{\dagger}(\vec{x}) \rightarrow
			 \psi^{\dagger}(\vec{x})e^{-i\vartheta}$
of the field operators, whereas their expectation values are clearly not.
Notice that the $U(1)$ symmetry of the Bose gas is closely related to the
conservation of particle number. This is most easily seen by observing
that the invariance of the Hamiltonian is due to the fact that each term in
the right-hand side of Eq.\ (\ref{hamil}) contains an equal number of
creation and annihilation operators. The relationship can also be
established in a more formal way by noting that the $U(1)$ gauge
transformations are generated by the particle number operator. As we will
see later on, it has important consequences for the dynamics of the order
parameter.

To understand why $\langle \psi(\vec{x}) \rangle$ is the order parameter
associated with Bose-Einstein condensation, it is convenient to use a
momentum-space description and to introduce the annihilation operator
for a particle with momentum $\hbar \vec{k}$
\begin{equation}
a_{\vec{k}} = \int d\vec{x} \: \psi(\vec{x})
		    \frac{e^{-i \vec{k} \cdot \vec{x}}}{\sqrt{V}}
\end{equation}
and the corresponding creation operator ${a_{\vec{k}}}^{\dagger}$ by
Hermitian conjugation. The basis of states for the gas is then characterized
by the occupation numbers $\{N_{\vec{k}}\}$. If the gas is condensed, there
is a macroscopic occupation of the zero-momentum state and the relevant
states are $|N_{\vec{0}}, {\{N_{\vec{k}}\}}_{\vec{k} \neq \vec{0}} \rangle$
with only $N_{\vec{0}}$ proportional to $N$. Within this subspace of states
we have
\begin{equation}
\langle {a_{\vec{0}}}^{\dagger} a_{\vec{0}} \rangle =
		  \langle N_{\vec{0}} \rangle \simeq
		  \langle N_{\vec{0}} \rangle + 1 =
		  \langle a_{\vec{0}} {a_{\vec{0}}}^{\dagger} \rangle
\end{equation}
and we can neglect that $a_{\vec{0}}$ and ${a_{\vec{0}}}^{\dagger}$ do not
commute. As a result we can treat these operators as complex numbers
\cite{bogol} and say that
$\langle {a_{\vec{0}}}^{\dagger} a_{\vec{0}} \rangle =
     \langle {a_{\vec{0}}}^{\dagger} \rangle \langle a_{\vec{0}} \rangle$
or equivalently that $\langle a_{\vec{0}} \rangle = \sqrt{N_{\vec{0}}}$.
In coordinate space the latter reads
$\langle \psi(\vec{x}) \rangle = \sqrt{n_{\vec{0}}}$, with
$n_{\vec{0}} = N_{\vec{0}}/V$ the condensate density.

The above argument essentially tells us that a sufficient condition for a
nonzero value of $\langle \psi(\vec{x}) \rangle$ is
$\langle N_{\vec{0}} \rangle \gg 1$. Although this is intuitively appealing,
it is important to point out that it is not generally true. Consider for
example the ideal Bose gas \cite{huang}. In the grand canonical ensemble the
total number of particles in the gas is given by
\begin{equation}
\label{density}
N = \sum_{\vec{k}} \langle N_{\vec{k}} \rangle =
    \sum_{\vec{k}} \frac{1}{\zeta^{-1} e^{\beta \epsilon_{\vec{k}}} - 1} \:,
\end{equation}
where $\beta$ is $1/k_{B}T$, $\epsilon_{\vec{k}}$ is the kinetic energy
$\hbar^{2} \vec{k}^{2}/2m$, $\zeta$ is the fugacity $e^{\beta \mu}$ and
$\mu$ is the chemical potential.

At high temperatures the fugacity is small and we are allowed to take the
continuum limit of Eq.\ (\ref{density}), which results in the equation of
state
\begin{equation}
\label{eqst}
n = \frac{1}{\Lambda^{3}} g_{3/2}(\zeta) \:,
\end{equation}
using the thermal de Broglie wavelength
$\Lambda = \sqrt{2 \pi \hbar^{2}/mk_{B}T}$ and the Bose functions
$g_{n}(\zeta)$ defined by
\begin{equation}
g_{n}(\zeta) = \frac{1}{\Gamma(n)} \int_{0}^{\infty} dx \:
				   \frac{x^{n-1}}{\zeta^{-1} e^{x} - 1} \:.
\end{equation}
Lowering the temperature while keeping the density fixed, the fugacity
increases until it ultimately reaches the value one at the critical
temperature
\begin{equation}
T_{0} = \frac{2 \pi \hbar^{2}}{mk_{B}}
		  {\left( \frac{n}{g_{3/2}(1)} \right)}^{2/3} \simeq
	\frac{2 \pi \hbar^{2}}{mk_{B}}
		  {\left( \frac{n}{2.612} \right)}^{2/3} \:.
\end{equation}
At this point Eq.\ (\ref{eqst}) ceases to be valid because the occupation
number of the zero-momentum state, which is equal to $\zeta/(1-\zeta)$,
diverges and must be taken out of the discrete sum in Eq.\ (\ref{density})
before we take the continuum limit. Moreover, we only need to treat the
zero-momentum term separately because in the thermodynamic limit the chemical
potential goes to zero as $V^{-1}$, whereas the kinetic energy for the
smallest nonzero momentum decreases only as $V^{-2/3}$. Consequently, below
the critical temperature the equation of state becomes
\begin{equation}
n = n_{\vec{0}} + \frac{1}{\Lambda^{3}} g_{3/2}(1)
\end{equation}
and leads to a condensate density equal to
\begin{equation}
\label{cond}
n_{\vec{0}} = n \left( 1- {\left( \frac{T}{T_{0}} \right)}^{3/2} \right) \:.
\end{equation}

We thus find that the average occupation number
$\langle N_{\vec{0}} \rangle$ is at all temperatures given by
$\zeta/(1-\zeta)$, i.e. its value in the grand canonical ensemble with the
density matrix $e^{-\beta(H - \mu N)}$. Since this density matrix
commutes with the particle number operator, we conclude that in the case of
an ideal Bose gas there is a macroscopic occupation of the zero-momentum
state without a spontaneous breaking of the $U(1)$ symmetry. To show
more rigorously that ${\langle \psi(\vec{x}) \rangle}_{eq} = 0$ at all
temperatures we determine the free-energy density of the gas as a function
of the order parameter $\langle \psi(\vec{x}) \rangle$. Dealing with a
noninteracting system it is not difficult to obtain
\begin{equation}
f(\langle \psi(\vec{x}) \rangle,T) =
		      - \mu(T) {| \langle \psi(\vec{x}) \rangle |}^{2}
\end{equation}
for a homogeneous value of the order parameter. Because $\mu \leq 0$ the
minimum is indeed always at $\langle \psi(\vec{x}) \rangle = 0$ and it is
necessary to identify the condensate density $n_{\vec{0}}$ with the order
parameter of the ideal Bose gas (cf.\ Eq.\ (\ref{cond})).

Notwithstanding the previous remarks, the order parameter for Bose-Einstein
condensation in a weakly-interacting Bose gas is given by
$\langle \psi(\vec{x}) \rangle$. This was put on a firm theoretical basis by
Hugenholtz and Pines \cite{hugen}, who calculated the free energy as a
function of the above order parameter and showed that at sufficiently low
temperatures the system develops an instability that is removed by a
nonzero value of $\langle \psi(\vec{x}) \rangle$. In addition, they derived
an exact relationship between the chemical potential and the condensate
density, which turns out to be valid also in the nonequilibrium problem of
interest here and is important for an understanding of how the $U(1)$
symmetry is broken dynamically.

\subsection{Condensation time}

We have argued that by means of evaporative cooling a doubly-polarized atomic
hydrogen gas can be quenched into the critical region of the phase transition
and that this kinetic part of the condensation process is described by a
quantum Boltzmann equation. As a result the gas acquires on the time scale
$\tau_{el}$ an equilibrium distribution with some temperature $T$, which is
slightly above the critical temperature $T_{0}$ of the ideal Bose
gas because a condensate cannot be formed at this stage.

For the study of the subsequent coherent part of the condensation process it
is therefore physically reasonable to assume that at a time $t_{0}$ the
density matrix $\rho(t_{0})$ of the gas is well approximated by the density
matrix of an ideal Bose gas with temperature $T$. The evolution of the order
parameter $\langle \psi(\vec{x}) \rangle$ for times larger than $t_{0}$ is
then completely determined by the Heisenberg equation of motion
\begin{equation}
i \hbar \frac{d\psi(\vec{x},t)}{dt} = [\psi(\vec{x},t) , H] \:,
\end{equation}
for the field operator. Substituting herein the Hamiltonian of
Eq.\ (\ref{hamil}) and taking the expectation value with respect to
$\rho(t_{0})$ we find
\begin{equation}
\label{heisen}
i \hbar \frac{d\langle \psi(\vec{x},t) \rangle}{dt} =
\frac{-\hbar^{2}\nabla^{2}}{2m} \langle \psi(\vec{x},t) \rangle +
   \int d\vec{x}' \: V(\vec{x} - \vec{x}')
	\langle \psi^{\dagger}(\vec{x}',t) \psi(\vec{x}',t)
					   \psi(\vec{x},t) \rangle \:,
\end{equation}
where the complicated part is of course the evaluation of
$\langle \psi^{\dagger}(\vec{x}',t)\psi(\vec{x}',t)\psi(\vec{x},t) \rangle$.
In lowest order we simply have
\begin{eqnarray}
\langle \psi^{\dagger}(\vec{x}',t) \psi(\vec{x}',t) \psi(\vec{x},t) \rangle
   \simeq \langle \psi^{\dagger}(\vec{x}',t) \rangle
	  \langle \psi(\vec{x}',t) \rangle \langle \psi(\vec{x},t) \rangle
      &+& \langle \psi^{\dagger}(\vec{x}',t) \psi(\vec{x}',t) \rangle
	  \langle \psi(\vec{x},t) \rangle   \nonumber \\
      &+& \langle \psi^{\dagger}(\vec{x}',t) \psi(\vec{x},t) \rangle
	  \langle \psi(\vec{x}',t) \rangle \:,
\end{eqnarray}
which after substitution into Eq.\ (\ref{heisen}) leads to
\begin{eqnarray}
\left( i \hbar \frac{d}{dt} + \frac{\hbar^{2}\nabla^{2}}{2m} \right)
   \langle \psi(\vec{x},t) \rangle
     =& & \int d\vec{x}' \: V(\vec{x} - \vec{x}')
	    \langle \psi^{\dagger}(\vec{x}',t) \psi(\vec{x}',t) \rangle
	    \langle \psi(\vec{x},t) \rangle   \nonumber \\
      &+& \int d\vec{x}' \: V(\vec{x} - \vec{x}')
	    \langle \psi^{\dagger}(\vec{x}',t) \psi(\vec{x},t) \rangle
	    \langle \psi(\vec{x}',t) \rangle  \nonumber \\
      &+& \int d\vec{x}' \: V(\vec{x} - \vec{x}')
	    \langle \psi^{\dagger}(\vec{x}',t) \rangle
	    \langle \psi(\vec{x}',t) \rangle
	    \langle \psi(\vec{x},t) \rangle \:,
\end{eqnarray}
and thus corresponds exactly to the Hartree-Fock approximation.

To proceed we must restrict ourselves to the case of a dilute Bose gas in the
quantum regime. Introducing the scattering length $a$, which is of the order
of the range of the interaction, the quantum regime is characterized by
$a/\Lambda \ll 1$. We therefore need to consider only $s$-wave scattering and
can neglect the momentum dependence of various collisional quantities. In
particular, we can replace the potential $V(\vec{x} - \vec{x}')$ by the
contact interaction $V_{\vec{0}} \, \delta(\vec{x} - \vec{x}')$ with
$V_{\vec{0}} = \int d\vec{x} \: V(\vec{x})$.
Hence, in the Hartree-Fock approximation we obtain
\begin{equation}
\left( i \hbar \frac{d}{dt} + \frac{\hbar^{2}\nabla^{2}}{2m} \right)
  \langle \psi(\vec{x},t) \rangle =
  \left( 2nV_{\vec{0}} +
	   V_{\vec{0}} {|\langle \psi(\vec{x},t) \rangle|}^{2} \right)
				       \langle \psi(\vec{x},t) \rangle \:,
\end{equation}
having only the trivial solution $\langle \psi(\vec{x},t) \rangle = 0$ for
a space and time-independent order parameter. Within this lowest order
approximation we thus conclude that $\tau_{nucl} = \infty$ and that
the formation of a condensate will not take place.

Fortunately, it is well known that the Hartree-Fock approximation is not
sufficiently accurate for a dilute Bose gas because the diluteness condition
$na^{3} \ll 1$ implies that we should consider all two-body processes, i.e.\
two particles must be allowed to interact also more than once. The
appropriate approximation is therefore the ladder or $T$-matrix approximation
and is diagrammatically explained in Fig.\ \ref{fig3}. Moreover, in the
degenerate regime where the temperature $T$ is slightly larger than $T_{0}$
and the degeneracy parameter $n\Lambda^{3}$ is of $O(1)$, the condition
$a/\Lambda \ll 1$ implies that also $na\Lambda^{2} \ll 1$ or physically
that the average kinetic energy of the gas is much larger than the
typical interaction energy. Consequently, an accurate discussion of the
nucleation of Bose-Einstein condensation in a weakly-interacting Bose gas
requires an evaluation of
$\langle \psi^{\dagger}(\vec{x}',t)\psi(\vec{x}',t)\psi(\vec{x},t) \rangle$
within the $T$-matrix approximation and in zeroth order in the gas
parameters $a/\Lambda$ and $na\Lambda^{2}$.

Although it is easy to formulate this objective, to actually perform the
calculation is considerably more difficult. It is most conveniently
accomplished by making use of the Keldysh formalism \cite{keldysh} which
has been reviewed by Danielewicz \cite{daniel} using operator methods. For a
functional formulation of this nonequilibrium theory and for the technical
details of the somewhat tedious mathematics we refer to our previous papers
\cite{stoof}. Here we only present the final results and concentrate on the
physics involved.

Due to the fact that we are allowed to neglect the (relative) momentum
dependence of the $T$ matrix, the equation of motion for the order parameter
$\langle \psi(\vec{x},t) \rangle$ acquires the local form of a
time-dependent Landau-Ginzburg theory
\begin{equation}
\label{motion}
\left( i \hbar \frac{d}{dt} + \frac{\hbar^{2}\nabla^{2}}{2m} \right)
  \langle \psi(\vec{x},t) \rangle =
  \left( S^{(+)}(t) +
	   T^{(+)} {|\langle \psi(\vec{x},t) \rangle|}^{2} \right)
				       \langle \psi(\vec{x},t) \rangle \:,
\end{equation}
which is recovered from a variational principle if we use the action
\begin{equation}
\label{action}
S(\langle \psi(\vec{x},t) \rangle,T) =
  \int dt \int d\vec{x} \:
      {\langle \psi(\vec{x},t) \rangle}^{*}
	  \left( i \hbar \frac{d}{dt}
	     + \frac{\hbar^{2}\nabla^{2}}{2m} - S^{(+)}(t)
	     - \frac{T^{(+)}}{2} {|\langle \psi(\vec{x},t) \rangle|}^{2}
	  \right)
       \langle \psi(\vec{x},t) \rangle \:.
\end{equation}
Here $S^{(+)}(t) \delta(t-t')$ is a good approximation for the retarded
self-energy $\hbar \Sigma^{(+)}(\vec{0};t,t')$ of a hydrogen atom with zero
momentum  and $T^{(+)} \simeq 4\pi\hbar^{2}a/m$ is the effective interaction
between two such atoms. Clearly, the action in Eq.\ (\ref{action}) is the
desired generalization of the Landau free energy and corresponds precisely to
the physical picture presented previously in Figs.\ \ref{fig1}(b) and
\ref{fig2}(b).

Therefore, $\tau_{nucl}$ is determined by the time dependence of
the coefficient $S^{(+)}$ which is shown in Fig.\ \ref{fig4} for three
different initial temperatures. If the temperature $T$ is much larger than
$T_{0}$, $S^{(+)}(t)$ is constant and equal to $8\pi\hbar^{2}an/m$. In this
region of the phase diagram coherent processes are negligible and the
evolution of the gas is described by a Boltzmann equation. Lowering the
temperature, the occupation numbers for momenta $\hbar k < O(\hbar/\Lambda)$
rise and lead to an enhancement of the coherent population of states with
momenta $\hbar k < O(\hbar \sqrt{na}) \ll O(\hbar/\Lambda)$. This is
signaled by the increasing correlation length
$\xi = \hbar/\sqrt{2mS^{(+)}(\infty)}$. At the critical temperature
$T_{c} = T_{0}( 1 + O(a/\Lambda_{0}) )$ we have $S^{(+)}(\infty) = 0$ and the
correlation length diverges. Below that temperature, but still above $T_{0}$
so as not to have a condensate already in the initial state, $S^{(+)}(t)$
actually changes sign and the gas develops the required instability for a
Bose-Einstein condensation. The change of sign takes place at
\begin{equation}
t \equiv t_{c} = t_{0} + O \left( \frac{a}{\Lambda_{c}}
				  \frac{\hbar}{k_{B}(T_{c}-T)} \right) \:,
\end{equation}
which shows that $\tau_{nucl}$ is in general of $O(\hbar/k_{B}T_{c})$ except
for temperatures very close to the critical temperature. Clearly, this time
scale is due to the fact that all states with momenta
$\hbar k < O(\hbar/\Lambda)$ cooperate in the coherent population of the
one-particle ground state.

After a small nucleus of $O(n(a/\Lambda_{c})^{2})$ has been formed, the
subsequent buildup of the condensate density is determined by the equation
of motion Eq.\ (\ref{motion}). Looking at the right-hand side we immediately
see that the time scale $\tau_{coh}$ involved in this process is typically
of $O(\hbar/n_{\vec{0}}T^{(+)})$ or equivalently of
$O((\hbar/k_{B}T_{c})(1/n_{\vec{0}}a\Lambda_{c}^{2}))$. Therefore,
$\tau_{coh} \gg \tau_{nucl}$ as anticipated in Fig.\ \ref{fig2}. The physical
reason for this time scale is that after the nucleation of the phase
transition the buildup of the condensate density is accompanied by a
depopulation of the momentum states with
$\hbar k < O(\hbar \sqrt{n_{\vec{0}}a})$. As a result it is not difficult to
show that in the limit $t \rightarrow \infty$ the condensate density is of
$O(na/\Lambda_{c})$ and thus that
$\tau_{coh} = O((\Lambda_{c}/a)^{2} \hbar/k_{B}T_{c})$.

Finally, it is interesting to point out how the gas can conserve the total
number of particles and apparently at the same time break the $U(1)$ gauge
symmetry that is responsible for this conservation law. To that end we write
the field operator $\psi(\vec{x},t)$ as the sum of its expectation value
$\langle \psi(\vec{x},t) \rangle$ and the fluctuation $\psi'(\vec{x},t)$,
and introduce a time-dependent chemical potential $\mu(t)$ by means of
\begin{equation}
\langle \psi(\vec{x},t) \rangle = \sqrt{n_{\vec{0}}(t)} \:
    exp \left( - \frac{i}{\hbar} \int_{t_{0}}^{t} dt' \: \mu(t') \right) \:.
\end{equation}
Substituting the latter in the action of Eq.\ (\ref{action}) and minimizing
with respect to $\sqrt{n_{\vec{0}}(t)}$ gives for $t > t_{c}$
\begin{equation}
n_{\vec{0}}(t) = \frac{- S^{(+)}(t) + \mu(t)}{T^{(+)}} \:,
\end{equation}
which determines the growth of the condensate density and is in effect a
nonequilibrium version of the Hugenholtz-Pines theorem \cite{hugen}.
Furthermore, by considering the fluctuations around $\sqrt{n_{\vec{0}}(t)}$
we can show that the chemical potential is determined by the constraint
\begin{equation}
n = n_{\vec{0}}(t) + \frac{1}{V} \int d\vec{x} \:
	  \langle \psi'^{\dagger}(\vec{x},t) \psi'(\vec{x},t) \rangle \:,
\end{equation}
enforcing the conservation of particle number at all times. In the complex
plane $\langle \psi(\vec{x},t) \rangle$ thus moves radially outward along a
spiral as shown in Fig.\ \ref{fig5}. Consequently, the phase of the order
parameter has never a fixed value and the $U(1)$ symmetry is not
really broken dynamically. This is of course expected since the system
evolves according to a symmetric Hamiltonian.

\section{CONCLUSIONS AND DISCUSSION}

We studied the evolution of a doubly-polarized atomic hydrogen gas in a
magnetic trap and showed that by means of evaporative cooling the gas can
accomplish the Bose-Einstein phase transition within its lifetime
$\tau_{inel}$. The condensation process proceeds under these conditions in
three stages: In the first kinetic stage the gas is quenched into the
critical region $T_{0} < T \leq T_{c}$. A typical time scale in this part of
the evolution is given by the time between elastic collisions $\tau_{el}$,
which for a degenerate gas is of $O((\Lambda_{c}/a)^{2} \hbar/k_{B}T_{c})$.
In the following coherent stage the actual nucleation takes place on the time
scale $\tau_{nucl} = O(\hbar/k_{B}T_{c})$ by means of a coherent population
of the zero-momentum state. The small nucleus formed in this manner then
grows on the much longer time scale
$\tau_{coh} = O((\Lambda_{c}/a)^{2} \hbar/k_{B}T_{c})$ by a depopulation of
the low-momentum states, having $\hbar k < O(\hbar \sqrt{n_{\vec{0}}a})$. In
the third and last stage of the evolution the Bogoliubov quasiparticles
produced in the previous stage have to come into equilibrium with the
condensate. This process can again be treated by a kinetic equation and was
studied by Eckern \cite{eckern}, who found that the corresponding relaxation
time $\tau_{rel}$ is of $O((\Lambda_{c}/a)^{3} \hbar/k_{B}T_{c})$. In the
case of atomic hydrogen this turns out to be comparable to the lifetime of
the system. Summarizing, we thus have the sequence
$\tau_{nucl} \ll \tau_{coh} \simeq
			    \tau_{el} \ll \tau_{rel} \simeq \tau_{inel}$
for the various time scales involved in the phase transition. The most
important requirement for the achievement of the phase transition is
therefore $\tau_{el} \ll \tau_{inel}$, which is relatively mild and should
not pose an insurmountable problem for future experiments aimed at the
realization of Bose-Einstein condensation.

Having arrived at this conclusion, it is necessary to discuss a recent paper
by Kagan, Svistunov, and Shlyapnikov \cite{kagan} that also considers the
evolution of a weakly-interacting Bose gas after the removal of the most
energetic atoms. In this paper the authors agree that the evolution of the
gas is divided into a kinetic and a subsequent coherent stage. Moreover,
their detailed study of the kinetic part of the evolution confirms our
conjecture that the gas is quenched into the critical region on the time
scale $\tau_{el}$. The investigation of the coherent part, however, leads to
the extreme result that a Bose-Einstein condensation cannot occur in a
finite amount of time. To understand why this conclusion is reached we
briefly present their line of thought.

At the end of the kinetic stage the gas has acquired large average
occupation numbers for the states with momenta
$\hbar k < \hbar k_{0} = O(\hbar \sqrt{n_{0}a})$, where $n_{0}$ is the
density of particles with these small momenta. Therefore, Kagan, Svistunov,
and Shlyapnikov argue that for a study of the coherent part of the evolution
we must use the initial condition
\begin{equation}
\label{initial}
\langle \psi(\vec{x},t_{0}) \rangle = \sum_{k<k_{0}}
	\sqrt{\langle N_{\vec{k}} \rangle} \:
	      \frac{e^{i \vec{k} \cdot \vec{x}}}{\sqrt{V}} \neq 0
\end{equation}
together with the nonlinear Schr\"{o}dinger equation
\begin{equation}
\label{schroe}
i \hbar \frac{d \langle \psi(\vec{x},t) \rangle}{dt} =
  \left( \frac{- \hbar^{2}\nabla^{2}}{2m} +
	    T^{(+)} {|\langle \psi(\vec{x},t) \rangle|}^{2} \right)
				       \langle \psi(\vec{x},t) \rangle \:,
\end{equation}
which has the equilibrium solution
$\langle \psi(\vec{x},t) \rangle = \sqrt{n_{0}} \, exp( -i \mu_{0} t )$ and
$\mu_{0} = n_{0}T^{(+)}$. Consequently, all the particles that have initially
momenta $\hbar k < \hbar k_{0}$ are in the limit $t \rightarrow \infty$
assumed to be in the condensate.

Linearizing the Hamiltonian of Eq.\ (\ref{schroe}) around this equilibrium
solution they then observe that the energy involved with a magnitude
fluctuation of the order parameter is $\epsilon_{\vec{k}} + n_{0}T^{(+)}$,
whereas the energy involved with a phase fluctuation is only
$\epsilon_{\vec{k}}$. As a result they assert that on the time scale
$\tau_{ampl} = \tau_{coh} = O(\hbar/n_{0}T^{(+)})$ a state is formed in which
the amplitude of $\langle \psi(\vec{x},t) \rangle$ is fixed, but the phase is
still strongly fluctuating because the corresponding time scale $\tau_{ph}$
is much longer and even diverges as $V^{2/3}$ in the thermodynamic limit.
Hence, for finite times the gas is in a state with a so-called
quasicondensate \cite{popov} and a real condensate is only formed in the
limit $t \rightarrow \infty$.

Clearly, this physical picture of two different time scales for the
amplitude and phase fluctuations of the order parameter is only applicable
if these fluctuations exist independently of each other. Looking only at the
Hamiltonian this indeed seems to be the case. However, a correct discussion
of the fluctuations must be based on the equations of motion or equivalently
the Lagrangian. The latter contains a first-order time derivative which
strongly couples the amplitude and phase fluctuations. Therefore, a dilute
Bose gas does not have two but only one dispersion relation, i.e.\ the
well-known Bogoliubov dispersion
$\sqrt{\epsilon_{\vec{k}}(\epsilon_{\vec{k}} + 2n_{0}T^{(+)})}$, and we are
lead to $\tau_{ph} = \tau_{ampl}$. It is interesting to note that in the case
of a neutral BCS-type superfluid we do have two different time scales because
the Lagrangian now contains a second-order time derivative and the amplitude
and phase fluctuations are indeed independent in lowest order
\cite{bog,anders}.

An even more serious problem with the approach of Kagan, Svistunov, and
Shlyapnikov is their claim that the use of the initial condition in
Eq.\ (\ref{initial}) is justified because
$\langle N_{\vec{k}} \rangle \gg 1$. As we have pointed out before this is
not true in general. For $\langle \psi(\vec{x},t) \rangle$ to be nonzero we
must show that the system has a corresponding instability. However, within
the $T$-matrix approximation it is not difficult to show that the instability
associated with a quasicondensate is always preceded by the instability
corresponding to the formation of a condensate. This implies that we always
have to take Bose-Einstein condensation into account first. After that has
been accomplished by means of the theory reviewed here, it is of course
no longer relevant to consider the appearance of a quasicondensate.

\section*{ACKNOWLEDGMENTS}

It is a great pleasure to thank Tony Leggett for various helpful discussions
and for giving me the opportunity to visit the University of Illinois at
Urbana-Champaign. I also benefited from conversations with Steve Girvin,
Daniel Loss, Kieran Mullen, and Jook Walraven. This work was supported by
the National Science Foundation through Grant No.\ DMR-8822688.


\begin{figure}
\caption{Visualization of (a) the time scale $\tau_{coh}$ for the relaxation
	 of the order parameter to its equilibrium value and (b) the
	 time scale $\tau_{nucl}$ associated with the appearance of the
	 instability.
	 \label{fig1}}
\end{figure}

\begin{figure}
\caption{Visualization of the time scales $\tau_{coh}$ and $\tau_{nucl}$,
	 using the time dependence of (a) the order parameter and (b) the
	 coefficient $\alpha$ of the quadratic term in the free energy.
	 \label{fig2}}
\end{figure}

\begin{figure}
\caption{Diagrammatic representation of the $T$-matrix equation. The wavy
	 line corresponds to the interaction and the straight line to
	 the non-interacting one-particle Green's function.
	 \label{fig3}}
\end{figure}

\begin{figure}
\caption{Time dependence of the coefficient $S^{(+)}$ for three
	 different initial temperatures of the Bose gas.
	 \label{fig4}}
\end{figure}

\begin{figure}
\caption{Evolution of the complex order parameter
	 $\langle \psi(\vec{x}) \rangle$, which is constrained by the
	 requirement of particle number conservation.
	 \label{fig5}}
\end{figure}

\end{document}